\documentclass[%
superscriptaddress,
preprint,
 amsmath,amssymb,
 aps,
prc
]{revtex4-1}

\usepackage{graphicx}
\usepackage{physics}
\usepackage{siunitx}

\begin{document}

\title{First-passage times in complex energy landscapes:\\
a case study with nonmuscle myosin II assembly}
\author{Rick~Bebon}
\author{Ulrich~S.~Schwarz}
\affiliation{Institute for Theoretical Physics, Heidelberg University, Philosophenweg 19, 69120 Heidelberg, Germany}
\affiliation{BioQuant, Heidelberg University, Im Neuenheimer Feld 267, 69120 Heidelberg, Germany}
\date{\today}

\begin{abstract}
Complex energy landscapes often arise in biological systems, e.g.~for protein folding, 
biochemical reactions, or intracellular transport processes. Their physical effects 
are frequently reflected in the first-passage times arising from these energy landscapes. 
However, their calculation is notoriously challenging and it is often difficult to identify the most relevant features of a
given energy landscape. Here we show how this can be achieved 
by coarse-graining the Fokker-Planck equation to a master
equation and decomposing its first-passage times in an iterative process. 
We apply this method to the electrostatic interaction between two
rods of nonmuscle myosin II (NM2), which is the main molecular motor for force
generation in nonmuscle cells. Energy landscapes are
computed directly from the amino acid sequences of the three 
different isoforms. Our approach allows us to identify 
the most relevant energy barriers for their
self-assembly into nonmuscle myosin II minifilaments
and how they change under force. In particular, we find that antiparallel configurations 
are more stable than parallel ones, but also show more changes under mechanical loading. 
Our work demonstrates the rich dynamics that can 
be expected for NM2-assemblies under mechanical load and in general
shows how one can identify the most relevant energy barriers
in complex energy landscapes.
\end{abstract}

\maketitle

\section{Introduction}
The concept of an energy landscape provides a powerful theory approach that has led to a unified description of many complex natural phenomena
in numerous different physical systems~\cite{wales2003energy}.
If depicted graphically, such systems resemble mountainous landscapes with many hills and valleys.
While notable examples from chemistry and condensed matter physics include chemical kinetics~\cite{espenson1995chemical,houston2012chemical}
or glass-forming systems~\cite{heuer2008exploring,garrahan2002geometrical,tuanase2003topological},
in a biophysical context, energy landscapes are central for explaining
a wide range of biomolecular processes~\cite{nussinov2014second}. 
For example, they help to understand the folding and conformational dynamics of
proteins~\cite{wales2003energy,wales2018exploring}, including single-molecule experiments~\cite{neupane2016protein,satija2020broad}, where details of the underlying 
multidimensional energy landscape can be inferred from transition paths of low-dimensional 
projections~\cite{satija2020broad}. They also 
provide perspectives on the structure and dynamics of the entire genome~\cite{zhang2017genomic}
and help us understanding the microscopic origins of selective transport processes through membrane 
channels~\cite{chou1999kinetics,bauer2010thermodynamics,berezhkovskii2002channel,zilman2009effects}.
Moreover, the general notion
of diffusion on energy landscapes has been proven to be successful
for e.g.~modeling molecular motors~\cite{julicher1997modeling,kolomeisky2013motor},
where motor-specific landscapes have been obtained from molecular structures~\cite{maisuradze2010relation,mukherjee2012realistic,mukherjee2017simulating} or using
single-molecule trajectories~\cite{hummer2001free,harris2007experimental,gupta2011experimental}.

Very often, the main aspect of interest for diffusive motion in a multi-well energy landscape
is the \emph{first-passage time} (FPT)~\cite{van1992stochastic,redner2001guide,metzler2014first},
i.e.\ the time when a stochastic process first passes a prescribed threshold value.
First-passage times tend to have a broad probability distribution, yet the average value
(\emph{mean first-passage time}, MFPT) often captures main physical aspects of complex systems.
Moreover, FPT-distributions and more so MFPTs
do not require a complete solution of the stochastic equations and therefore tend to be more accessible. 
Many biophysical systems have been successfully analyzed along these lines, 
including the kinetics of chemical 
reactions~\cite{kramers1940brownian,schulten1981dynamics,hanggi1990reaction,szabo1980first,benichou2010geometry}, 
enzyme catalysis~\cite{rotbart2015michaelis,singh2017statistical,robin2018single},
molecular search processes~\cite{condamin2007first,eliazar2007searching,benichou2011intermittent,chou2014first,iyer2016first},
signaling in biological cells~\cite{benichou2014first,godec2016first,godec2016universal,grebenkov2018towards,grebenkov2016universal},
cell adhesion~\cite{erdmann2004stochastic,erdmann2007impact,blom2021criticality,schwarz2006focal,erdmann2004stability},
modeling of molecular motors~\cite{erdmann2013stochastic,erdmann2012stochastic},
self-assembly processes~\cite{baschek2012stochastic},
protein folding kinetics~\cite{chekmarev2021first,micheelsen2003mean}, and
single-molecule pulling experiments by atomic force spectroscopy~\cite{shillcock1998escape,hummer2003kinetics,dudko2006intrinsic,dudko2007extracting,dudko2008theory,hu2010first,abkenar_dissociation_2017}.
More broadly, first-passage theory is central to the description of diverse problems ranging from
foraging behavior of animals or bacteria~\cite{viswanathan1996levy,viswanathan2011physics,bell2012searching,palyulin2014levy,berg2018random},
population dynamics~\cite{johnson1992animal,chou2014first},
and the spread of infectious diseases~\cite{lloyd2001viruses,volovik2010first,hufnagel2004forecast},
to transport in disordered media~\cite{ben2000diffusion}, 
or the dynamics of financial markets and stock options~\cite{mantegna1999introduction}.

While substantial advances have been achieved in the theory of first-passage processes on flat energy landscapes 
for finite domains~\cite{condamin2007first,holcman,benichou2010geometry},
much less is known about first-passage properties for more general rough energy landscapes~\cite{zwanzig1988diffusion}.
Usually such results are restricted to simple model landscapes and do not address the complex
energy landscapes typically encountered in biological systems. 
For instance, it was found that the introduction of finite potential barriers may lead to a reduction in the MFPT~\cite{palyulin2012finite} which subsequently allows for the optimization of escape rates~\cite{chupeau2020optimizing},
and that intermediates barriers may accelerate the rate of barrier crossing events~\cite{wagner1999intermediates}.
Moreover, it has been shown that first-passage time distributions provide a link between the dynamical properties and the topology of complex networks~\cite{li_mechanisms_2013}. 
For one-dimensional discrete-state dynamics these theoretical results made it possible to quantitatively infer key features of energy landscapes from measurements of the first-passage time distribution.
A characteristic power-law regime for the short-time statistics of the first-passage time was demonstrated to reflect e.g.~the number and depth of intermediate potential minima in experiments on different length scales~\cite{thorneywork2020direct}.
Furthermore, a recently discovered duality between first-passage time and relaxation processes for reversible
Markovian dynamics provided further insight into first-passage time phenomena in rugged energy landscapes~\cite{hartich2018duality,hartich2019interlacing}. Yet a more general
understanding of the first-passage properties for general rough landscapes is still missing.

In this paper, we use a specific biophysical example to demonstrate how in general the MFPT can be calculated for highly complex and rugged 
one-dimensional energy landscapes by coarse-graining the dynamics following the continuous overdamped Fokker-Planck equation
into a discrete master equation. As our case study, we use self-assembly of the 
molecular motor nonmuscle myosin II (NM2). 
Myosins constitute a diverse superfamily of motor proteins that
generate mechanical work by converting chemical energy produced during ATP-hydrolysis into conformational changes that eventually propel actin filaments forward~\cite{hartman2012myosin,alberts2017molecular}.
The myosin II subfamily is a class of non-processive molecular motor proteins that have to work in sufficiently large
ensembles to generate force. This is achieved by self-assembly into bipolar filaments,
which are present not only in muscle cells, but also in nonmuscle cells, where NM2
forms so-called minifilaments that are composed of up to 30 NM2 molecules
(each with two motor heads) with a typical size of 
\SI{300}{\nm}~\cite{craig2006structure,vicente2009non,dasbiswas2018ordering}.
NM2 is the main force generator in nonmuscle cells and crucial for many
essential cellular functions, including adhesion, migration, division, and mechanosensing.
While NM2 traditionally has been investigated mainly using electron microscopy~\cite{niederman1975human,billington2013characterization},
super-resolution microscopy has made it possible 
to break the \SI{200}{\nm} resolution limit of traditional light microscopy,
allowing the investigation of NM2 minifilaments dynamics in 
live cells~\cite{burnette2014contractile,beach2014nonmuscle,fenix2016expansion,hu2017long,jiu2019myosin}.
These live-cell imaging studies demonstrated that besides \emph{de novo} filament formation, the proliferation of NM2 minifilaments is 
additionally driven by novel partitioning mechanisms like filament splitting and reveal unexpected processes such as expansion, 
concatenation, long-range attraction and stacking.
Adding to the notion that NM2 minifilamens are far more dynamic than formerly appreciated, it was observed
that individual NM2 monomers are characterized by high exchange rates~\cite{hu2017long} and that the three human NM2-variants (A,B,C)
co-assemble into mixed filaments with distinct characteristics~\cite{beach2015myosin,weissenbruch2021distinct,weissenbruch2022}. 
Thus the question arises how the 
energy landscapes of these assemblies look like and how the system explores its different configurations.

NM2-molecules assemble by forming parallel and antiparallel staggers of their rod-like regions.
This interaction is mainly of electrostatic origin and determined by the amino acid sequence of the respective
rods~\cite{kaufmann2020electrostatic,ricketson2010multiple,straussman2005skip,mclachlan1982periodic,rosenberg2008mhc}.
Earlier modelling approaches were able to predict the experimentally observed staggers as local minima 
of the electrostatic interaction energy landscape, but also led to very noisy data~\cite{straussman2005skip,mclachlan1982periodic,rosenberg2008mhc}. 
A detailed three-dimensional molecular model of the NM2 rod structure provided
better data quality of the energy landscape, but again identified the known staggers
only as local minima~\cite{ricketson2010multiple}. 
Recently, the identification
of staggers were improved considerably by introducing the concept of MFPTs into
the modelling of NM2-staggering~\cite{kaufmann2020electrostatic}. 
Modeling in addition the splaying of NM2-molecules in a rolling and zipping motion
as part of the energy landscape, the
experimentally observed staggers emerged as maxima in the MFPT (e.g.~contact time) as
function of the initial alignment. Thus by replacing energies by MFPTs, the
physical stability of these systems became more apparent.

Although resulting in many interesting predictions on the stability of competing staggers, the 
earlier work did not identify the decisive features of the energy landscape that lead to these staggers.
It also did not consider the effect of mechanical force, that is essential for 
the functioning of NM2-minifilaments and that might stabilize new configurations or 
completely destabilize the assembly. Here we will use systematic and rigorous 
coarse-graining of the complex energy landscapes to address these important questions for
the assembly of NM2-molecules. In order to be able to
conduct a comprehensive analysis of a one-dimensional model, here we consider only 
straight myosin II rods and disregard the effects of bending.
Our work does not only give new and interesting
results for NM2-assembly, but also demonstrates how a successful coarse-graining procedure
can be implemented for complex one-dimensional energy landscapes in general.

\section{Derivation of the energy landscapes}

\subsection{Structural model and charge distribution of NM2-rod domains}
\label{sec:charges}

\begin{figure}[t]
	\includegraphics[width=\textwidth]{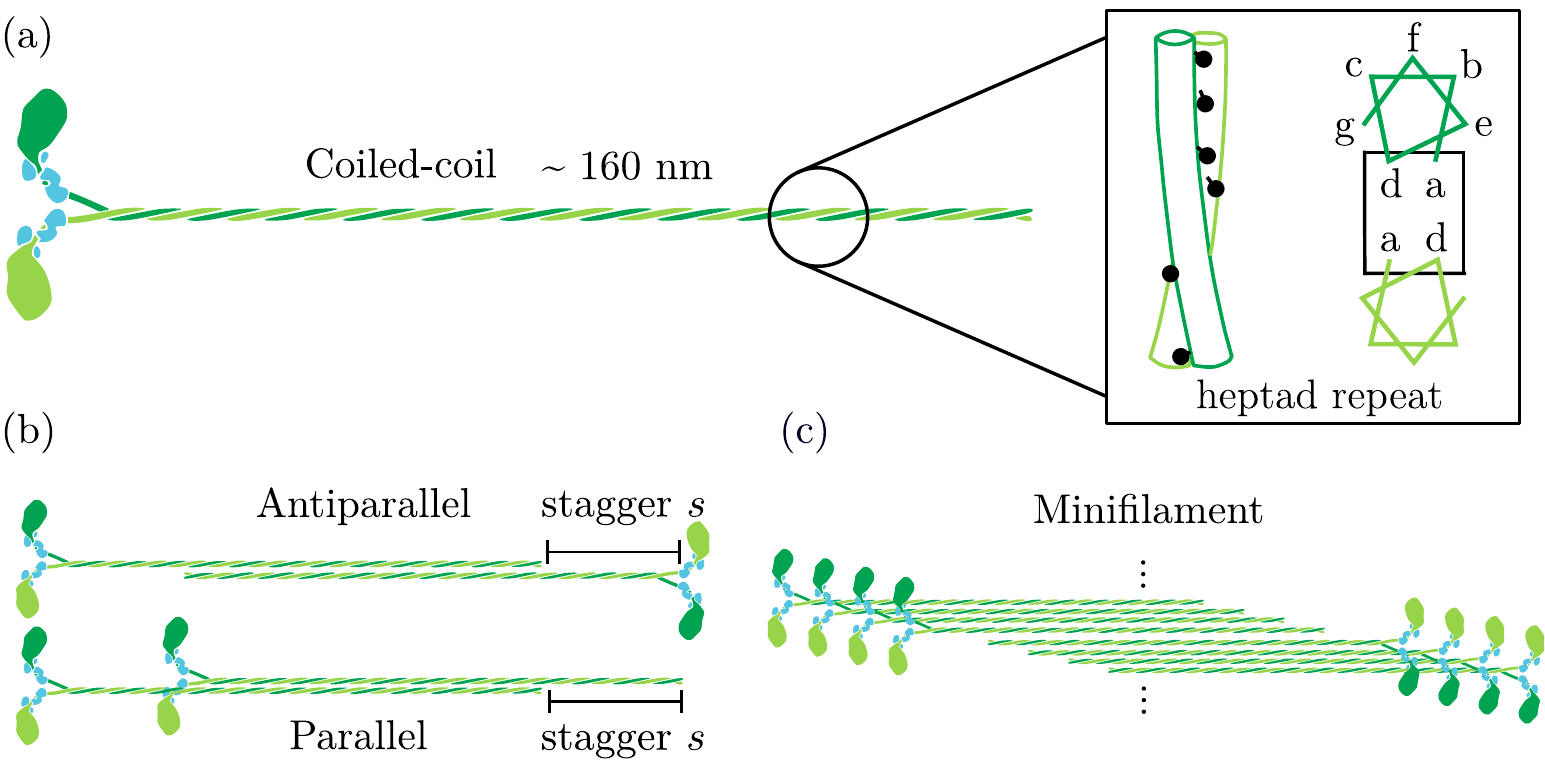}
	\caption{Nonmuscle myosin II (NM2) structure and assembly.
	(a) A NM2-molecule is mainly comprised of two identical heavy chains (light and dark green) that begin with a 
	motor head domain at the N-terminal domain and end with the nonhelical tail at the C-terminal.
	In addition, four light chains attach at the neck (blue). The rod part of the
	molecule consists of a \SI{160}{nm} long coiled-coil region in which the heavy chains
	wind around each other with a pitch of 3.5 residues per turn.
	The coiled-coil is stabilized by the heptad repeat ($a$ to $g$) of the amino 
	acid sequence of each heavy chain (inset).
	Residues at position $a$ and $d$ form a hydrophobic interface and the remaining residues are exposed 
	to the surrounding.
	(b) The electrostatic interactions of the exposed residues of the heptad repeats lead
	to either parallel or antiparallel staggers of the NM2-dimers.
	(c) Self-assembly of up to 30 NM2-molecules forms the bipolar minifilaments 
	with a typical size of \SI{300}{\nm}. 
	}
	\label{fig1}
\end{figure}

In order to give an instructive example of a complex one-dimensional energy landscape 
with many potential barriers of different height, we first introduce
the case of the self-assembly of NM2-molecules into minifilaments.
As schematically depicted in Fig.~\ref{fig1}a, one NM2 monomer consists mainly of two identical
myosin II heavy chains (shown in light and dark green) that dimerize to form its rigid helical backbone.
Hereby, each heavy chain begins with a globular motor head domain at the N-terminal containing 
bindings sites for actin filaments as well as for ATP-molecules. Following the head region comes a neck
region in which two regulatory light chains and two essential light chains bind (shown in blue).
Then comes the roughly \SI{160}{\nm} long and relatively rigid coiled-coil rod
and finally the non-helical tailpiece that disrupts the coiled-coil motif.
Although the whole complex is in fact a hexamer, here we call it a NM2-molecule for simplicity.
In the coiled-coil rod domain the two heavy chains wind around each other with
an axial spacing of \SI{0.1456}{\nm} between adjacent residues and 
a pitch of $3.5$ residues per turn due to interactions between periodically placed hydrophobic residues~\cite{lupas1996coiled,squire1971general}.
For this purpose, each NM2 heavy chain features an amino acid sequence with a heptad repeat
labeled from $a$ to $g$ as schematically depicted in the inset of Fig.~\ref{fig1}a.
Here, residues at positions $a$ and $d$ form the hydrophobic interface between two adjacent side chains, 
whereas the remaining residues are hydrophilic and exposed to the solvent.
The assembly of the NM2-molecules into dimers is mediated by the electrostatic
interactions of the exposed residues, whose charge pattern leads to either
parallel or antiparallel staggers, as shown in Fig.~\ref{fig1}b. 
The predominance of electrostatic driving forces during the assembly process was 
demonstrated by \emph{in vitro} experiments where an increase
in salt concentration was found to inhibit filament formation due to electrostatic screening effects~\cite{nakasawa2005critical}. The final minifilament contains up to 
30 NM2-molecules, is bipolar and around \SI{300}{\nm} large, compare Fig.~\ref{fig1}c.

\begin{figure}[t]
	\centering
	\includegraphics{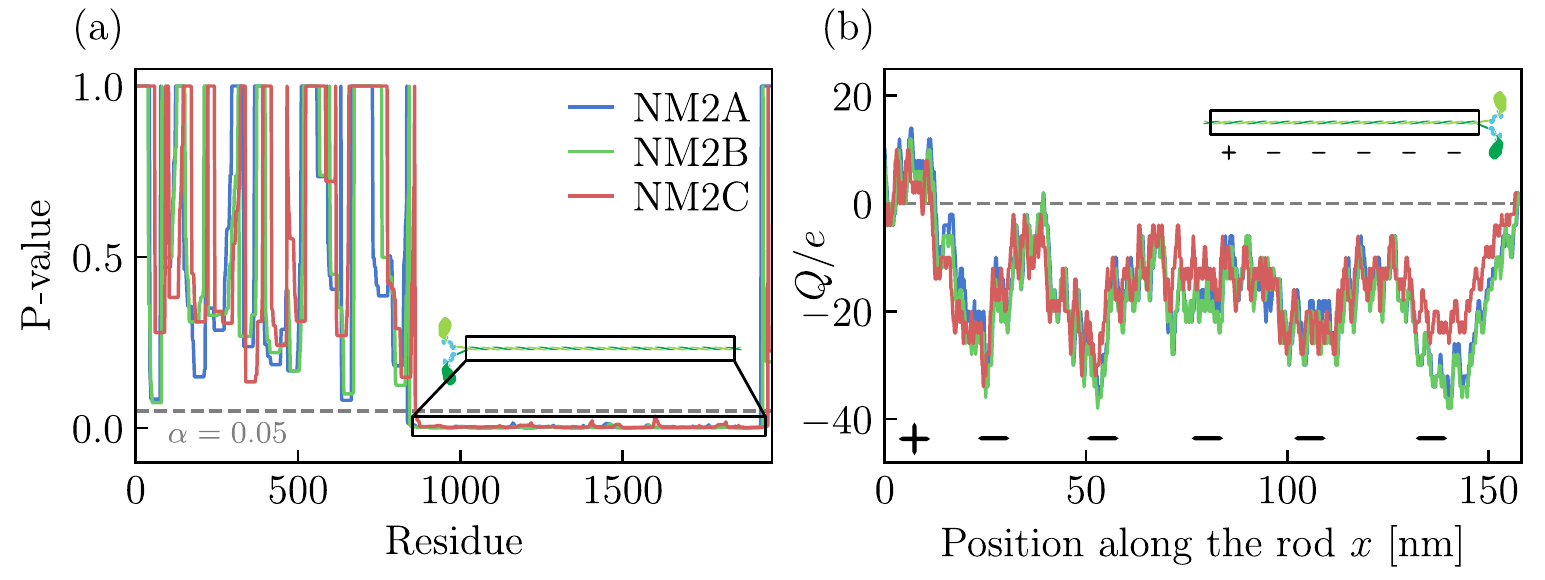}
	\caption{Analysis of NM2 amino acid sequences. 
	(a) Prediction of the coiled-coil region for different NM2-isoforms obtained with
	the software package \emph{Paircoil2}~\cite{mcdonnell2006paircoil2}.
	The dashed black line denotes the significance level of $\alpha = 0.05$ indicating
	that residues from roughly $850$ to $1920$ are part of the coiled-coil motif.
	Remaining residues on the left and right correspond to the head and
	to the non-helical tailpiece, respectively. 
	(b) Charge distribution along the NM2-rod treated as a chain of linear charges.
	Summation with a sliding window of $98$ residues highlights the expected positively charged C-terminal tip which is followed by five negatively charged regions along the whole rod domain. 
	}
	\label{fig2}
\end{figure}

The derivation of the electrostatic interaction energy between two NM2-molecules relies on the 
procedures recently introduced in Ref.~\cite{kaufmann2020electrostatic}.
First the amino acid sequence of the coiled-coil rod domain of individual NM2-molecules,
as provided by the \emph{NCBI Protein Database}~\cite{geer2010},
is translated into a linear chain of point-charges. We have
analyzed the amino acid sequences of the three human NM2-isoforms 
(denoted as NM2A, NM2B and NM2C) via the sequence-based coiled-coil prediction software 
\emph{Paircoil2}~\cite{mcdonnell2006paircoil2}.
By employing pairwise residue probabilities~\cite{berger1995predicting}, \emph{Paircoil2} is able to identify the location of coiled-coil motifs in 
protein amino acid sequence data.
Performing this statistical analysis yields a p-value for each individual residue of the sequence data, 
which in turn allows us to test the
null hypothesis whether the respective residue is \emph{not} located within a coiled-coil domain.
As depicted in Fig.~\ref{fig2}a, this hypothesis is \emph{rejected} for a domain ranging from roughly the 850th up to
the 1920th amino acid residue for all NM2-variants, indicating that these residues 
are indeed part of the coiled-coil motif 
of the NM2 heavy chain.
The first large non coiled-coil region indicates the motor and neck region, whereas 
the small region at the end of the amino sequence represents the non-helical tailpiece (compare Fig.~\ref{fig1}a).

After identifying the NM2 coiled-coil motif ($\sim 1100$ out of total $\sim 2000$ amino acids) 
and the positions of individual residues along the rod, 
we are now able to associate all required amino acids of the coiled-coil with their corresponding charges.
Amino acids at residue position $a$ and $d$ are hereby assigned zero net charge as they face inwards and
form the \emph{hydrophobic} coiled-coil backbone (compare inset in Fig.~\ref{fig1}a).
The remaining residue positions are either assigned a charge of ``$+2e$'' for positively charged amino acids (i.e.~arginine and lysine)
or ``$-2e$'' for negatively charged amino acids (i.e.~glutamic acid and aspartic acid).
Moreover point-like charges are arranged with an axial spacing of 
$\Delta x = \SI{0.1456}{\nm}$ between adjacent 
neighbors~\cite{lupas1996coiled,squire1971general,kaufmann2020electrostatic}
which subsequently yields the correct dimensions for the NM2-molecule length of approximately $\SI{160}{\nm}$.
The corresponding distribution of charges along the NM2 coiled-coil rod domain
is depicted in Fig.~\ref{fig2}b for all three different NM2-variants.
For clarity, we hereby smooth the distribution of charges via a sliding window of $98$ 
residues~\cite{straussman2005skip,kaufmann2020electrostatic} -- the preferred axial shift between interacting 
molecules due to the 196 residue repeat -- which allows us
to identify the positive C-terminal tip (Assembly Critical Domain, ACD) and the five regions with negative charge
spaced along the rod domain.
We note that the quantitative difference between the different NM2-variants is comparatively small.

\subsection{Electrostatic interaction}
\label{sec:elec_int}

The + - - - - - charge pattern revealed in Fig.~\ref{fig2}b is the physical basis
of the self-assembly of minifilaments and has been identified before by similar procedures~\cite{ricketson2010multiple,straussman2005skip,kaufmann2020electrostatic}.
In line with these previous modelling approaches,
our calculations of energy landscapes are restricted to the electrostatic interactions of two aligned NM2-molecules which are based on the charge distribution calculated in Sec.~\ref{sec:charges}.
For simplicity we do not account for bending of NM2 rods that has been used before by including bending energies
to complement the electrostatic interactions~\cite{kaufmann2020electrostatic} as it would require additional numerical procedures and effectively renders the energy landscapes two-dimensional.
We also neglect potential contributions of the entropy
of segment bending to the free energy. Because the persistence length
of the myosin II rods is \SI{130}{\nm}~\cite{hvidt_flexibility_1982,hvidt_flexibility_1983} 
and dangling parts to the polymers are
usually in the same order of magnitude or less, such contributions
are expected to contribute only little in our context.
To calculate the electrostatic interaction energy between two rods, 
we note that under physiological salt concentrations we are in the 
weak coupling regime for a rod with
$\SI{0.085}{e}$ per residue, 
a distance \SI{0.1456}{\nm} between residues and a radius \SI{1}{\nm}~\cite{kaufmann2020electrostatic}.
Thus the appropriate treatment of electrostatics is 
Poisson-Boltzmann theory~\cite{israelachvili2011intermolecular,poon2006soft}.
Our considered system is additionally characterized by small ionic strengths, allowing us to linearize
the Poisson-Boltzmann equation, yielding the Debye-Hückel equation 
\begin{equation}
	\nabla^2 \psi(\mathbf{r}) = \kappa^2 \psi(\mathbf{r}).
	\label{eq:dhe}
\end{equation}
Here $\psi$ denotes the electrostatic potential and $\kappa$ is the inverse Debye-Hückel screening length that
sets the typical length scale of the electrostatic screening.

To now compute the electrostatic interaction energy of two aligned NM2 rods, each modeled as a linear chain of charges,
we always have to consider two different cases.
As schematically illustrated in Fig.~\ref{fig1}b, NM2-molecules assemble in either \emph{parallel} or \emph{antiparallel} fashion, and they may align with a different stagger $s$.
For the geometry of a single point-particle with charge $q$, it is possible to analytically solve Eq.~\eqref{eq:dhe}, resulting
in a Yukawa-type potential~\cite{israelachvili2011intermolecular} of the form
\begin{equation}
    \psi (r) =  \frac{q}{4 \pi \epsilon \epsilon_0} \frac{\exp(-\kappa r)}{r}. 
\end{equation}
Exploiting the linearity of the Debye-Hückel equation~\eqref{eq:dhe} furthermore allows us to directly compute the total electrostatic interaction
energy $U(s)$ of two aligned NM2-rods as a function of their stagger $s$ as
\begin{equation}
	U(s) = \sum_{i=1}^N \sum_{j=1}^M \frac{q_i q_j}{4 \pi \epsilon \epsilon_0} \frac{\exp(-\kappa r_{ij})}{r_{ij}},
	\label{eq:landscape}
\end{equation}
where we define $r_{ij} \equiv \sqrt{y^2 + [(i-j)\Delta x -s ]^2}$ as the distance between residues. 
Point charges are denoted with $q$ whereas $N$ and $M$ denote the total number of residues of the respective rods with $i$ and $j$ as their corresponding index.
The distance along the NM2 rod, i.e.~the main axis, is given by $x$, while $y$ denotes the lateral spacing between two aligned NM2-molecules.
In the case of an antiparallel NM2-dimer configuration we invert one of the two charge sequences.

\begin{figure}[t]
	\centering
	\includegraphics{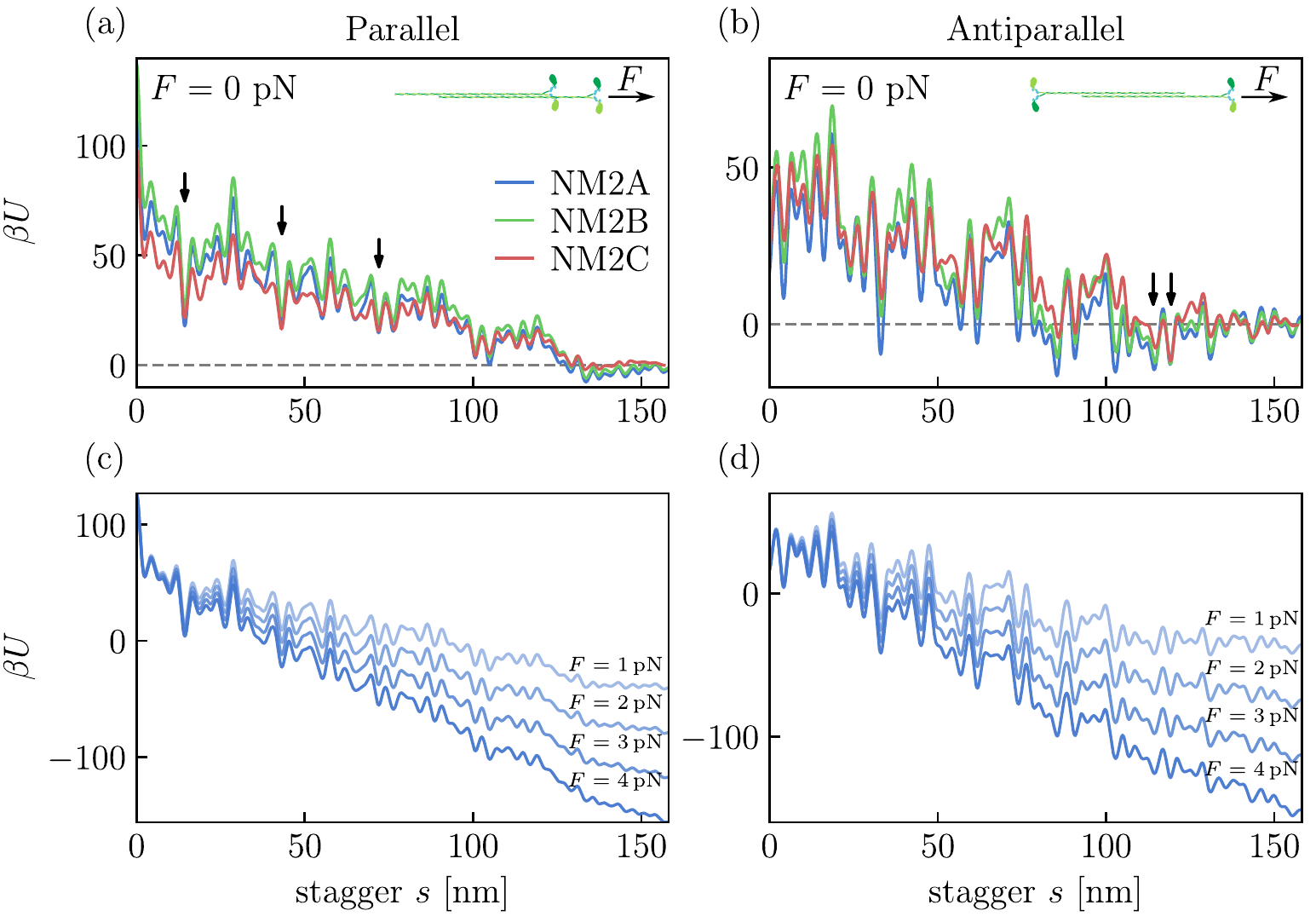}
	\caption{Energy landscapes of the electrostatic interaction for two 
	NM2-molecules for variable pulling force $F$.
	Landscapes for all three NM2 variants without external pulling in either (a) parallel or 
	(b) antiparallel configurations and as a function of their respective stagger $s$ 
	in units of the thermal energy $k_{\rm B}T = 1/\beta$.
    NM2-monomers are aligned with a lateral distance of $y=\SI{2}{\nm}$.
    Experimentally known staggers (black arrows) are local minima in the energy landscapes.
    Application of an variable external pulling force $F>\SI{0}{pN}$ tilts the (c) parallel and 
    (d) antiparallel energy landscape as explicitly shown for the NM2A-dimer.}
	\label{fig3}
\end{figure}

The resulting electrostatic interaction energy landscapes for NM2-dimers are shown
in Fig.~\ref{fig3}a and b as a function of the stagger $s$ and in units of the thermal energy $k_{\rm B}T = 1/\beta$
for parallel and antiparallel configurations, respectively. 
By calculating the interaction energy according to Eq.\eqref{eq:landscape} we assume a lateral spacing
of $y = \SI{2}{\nm}$~\cite{niederman1975human,kaufmann2020electrostatic}.
Moreover, we do not take negative staggers into account as their positive
ACDs would not be in contact with the negative charged region, i.e.~their interaction would become highly unfavorable.
The experimentally known staggers ($s=\SI{14.3}{\nm}$, $s=\SI{43.2}{\nm}$ and $s=\SI{72.0}{\nm}$
for parallel alignment and $s=\SI{113}{\nm}$ to $s=\SI{118}{\nm}$ for
antiparallel alignment) are marked in Fig.~\ref{fig3}a and b by black arrows and emerge
as local minima generated by the electrostatic interaction between
the positive ACD and regions
with increased net negativity along the NM2-rod. However, as noted earlier, 
these minima are not very clear, which motivates our approach to
use these energy landscapes to investigate the corresponding FPT-problems.

To investigate situations in which the NM2-dimer is being pulled apart, e.g.~due to contractile forces in the actomyosin network, we study the impact of an external pulling force $F$ on the energy landscape.
Force-mediated rearrangements may be of particular importance at the beginning stages of minifilament assembly and during the dynamical partitioning of minifilaments~\cite{fenix2016expansion, beach2017actin},
when filaments physically split into a pair of two ``daughter'' filaments. 
To account for the mechanical load we apply an additional constant external pulling 
force $F > \SI{0}{pN}$ along the stagger main-axis coordinate in positive $s$-direction.
As seen in Figs.~\ref{fig3}c and~\ref{fig3}d for parallel and antiparallel 
alignment, respectively, this driving leads to a tilted
NM2-dimer energy landscape
\begin{equation}
	\beta U^{(F)}(s) = \beta U_0(s) - \beta F s, 
	\label{eq:tilted_landscape}
\end{equation}
where $U_0 \equiv U^{(\SI{0}{pN})}(s)$ denotes the electrostatic potential energy as given by Eq.~\eqref{eq:landscape} of the original ``untilted'' NM2-dimer.

\section{Mean-first passage time until detachment}
\label{sec:MFPT_Cont}

In order to better identify relevant staggers during NM2-assembly,
we next compute the average time that passes until a NM2-dimer detaches from a certain initial value for the stagger.
In the biologically relevant overdamped limit, 
the probability density $p(s,t)$ to find the system at some time $t$ with a stagger $s$ evolves according
to the Fokker-Planck equation
\begin{equation}
	\partial_t p(s,t) = \mathcal{L} p(s,t)
	\label{eq:fpe}
\end{equation}
where $\mathcal{L} = -\partial_s A(s,t) + \partial_s^2 B(s,t)$ denotes the Fokker-Planck operator.
The drift term is determined by the external potential in
the form of $A = -\mu \partial_s U(s)$ where $\mu$ is the mobility coefficient (i.e.~inverse friction coefficient).
Moreover, the stochastic contribution of the Brownian motion is described by the diffusion constant via the Einstein relation 
$B = D = \mu / \beta $, where $\beta$ denotes the inverse thermal energy. 

We define the one-dimensional domain of interest $[a,b]$ of the detachment process as follows.
The absorbing boundary $b$ is set to the maximal possible stagger (or minimal overlap) between two aligned NM2 rods,
i.e., where the stagger is equal to
the complete rod length such that $b = 1085\Delta x \approx \SI{157.98}{\nm}$ for rods with 1085 amino acid residues.
Reaching the absorbing target therefore corresponds to the detachment of the NM2-dimer and marks the end of the process.
The reflecting boundary $a$ is placed at the minimal stagger (or maximal overlap) with
$a = \SI{0}{\nm}$, which corresponds to the complete alignment with the maximal overlap of the complete NM2 rod length.
Since the systems behaviour is described by absorbing and reflecting boundaries, our
FPT-problem is characterized by mixed boundaries with $a < b$.

\begin{figure}[t]
	\includegraphics{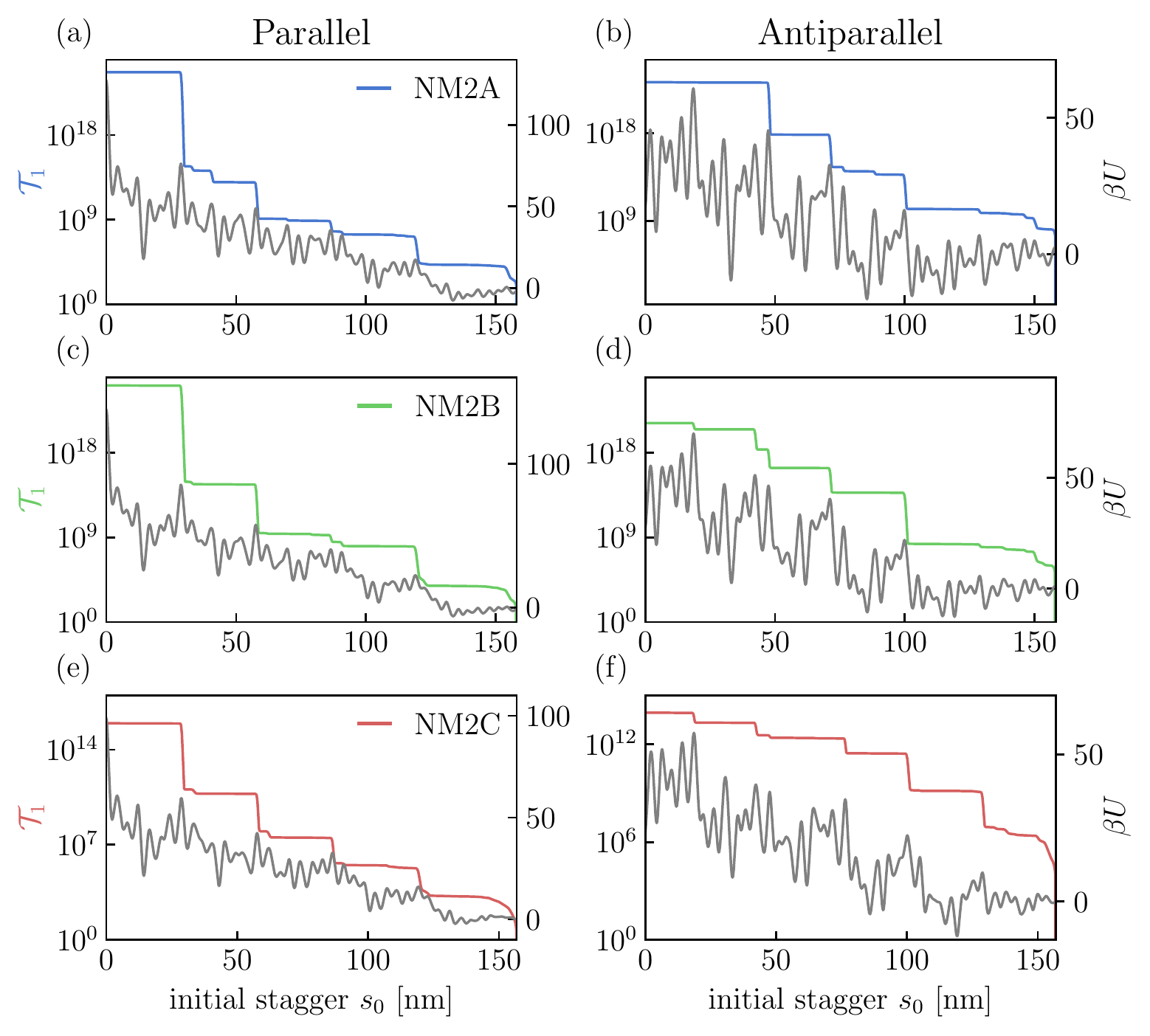}
	\caption{Mean first-passage time (MFPT) $\mathcal{T}_1$ calculated numerically
	from Eq.~\eqref{eq:mftp_dimer_cont} as a function of the initial stagger 
	position $s_0$ in either parallel (left) or antiparallel (right) configurations.
	Shown are the variants NM2A (blue, a+b), NM2B (green, c+d) and NM2C (red, e+f).
	In all cases the MFPT decreases monotonically with increasing $s_0$, however, $\mathcal{T}_1$ 
	shows significant decreases at only a few specific positions that coincide with certain 
	potential barriers of the underlying energy landscapes.
	While $\mathcal{T}_1$ for NM2A and NM2B are very similar, the one for NM2C is much smaller.
	}
	\label{fig4}
\end{figure}

The backward Fokker-Planck framework introduces an efficient way of calculating
the $n$-th moment $\mathcal{T}_n$ of the first-passage time density (FPTD) by employing the
recurrence relation~\cite{gardiner1985handbook,risken1996fokker,pavliotis2014stochastic}
\begin{equation}
	\mathcal{L}^\dagger \mathcal{T}_n = -n \mathcal{T}_{n-1} 
	\label{eq:recurrence_relation}
\end{equation}
with $\mathcal{T}_0 = 1$ and where 
$\mathcal{L}^\dagger = A(s_0,t)\partial_{s_0} + B(s_0,t)\partial_{s_0}^2$ denotes the backward
Fokker-Planck operator.
For a diffusion process subject to the NM2 energy landscape Eq.~\eqref{eq:landscape}, the mean first-passage time (MFPT) ($n=1$) consequently can be obtained by solving
the ordinary differential equation 
\begin{equation}
	\mathcal{L}^\dagger \mathcal{T}_1(s_0) = -1\ .
	\label{eq:bfpe_ode}
\end{equation}
In our case of mixed boundary conditions with $a < b$,
the solution of Eq.~\eqref{eq:bfpe_ode}, i.e.~the MFPT  $\mathcal{T}_1$,
is given by the well known formula~\cite{gardiner1985handbook,risken1996fokker,pavliotis2014stochastic}
\begin{equation}
	\mathcal{T}_1 (s_0) = \int_{s_0}^b\frac{dy}{\Psi(y)} \int_a^y dz \ \frac{\Psi(z)}{B(y)}
\end{equation}
where we define
\begin{equation}
	\Psi(z) \equiv \exp \left(\int^z dx \ \frac{A(x)}{B(x)} \right) = \exp (- \beta U(z)) 
\end{equation}
by inserting the drift velocity $A$ and diffusion coefficient $B$ as they enter the Fokker-Planck equation~\eqref{eq:fpe}.
The final analytical expression for the MFTP time now reads
\begin{equation}
	\mathcal{T}_1(s_0) = \frac{1}{D} \int_{s_0}^b dy \ e^{-\beta U(y)} \int_a^y dz \ e^{\beta U(z)}.
	\label{eq:mftp_dimer_cont}
\end{equation}
To calculate the MFTP for landscapes of different NM2-dimer variants we numerically solve 
Eq.~\eqref{eq:mftp_dimer_cont} after inserting the corresponding 
electrostatic interaction energy $U(s)$ from Eq.~\eqref{eq:landscape} for different NM2-variants and configuration.
Here the two integrals are solved numerically using the trapezoid rule function \texttt{integrate.trapezoid} as implemented in the python library \texttt{SciPy}~\cite{2020SciPy-NMeth}.
Throughout the remainder of this paper, we make the diffusion coefficient dimensionless, $D=1$, since we are interested in
the relative differences of the average detachment time between all NM2-variants and configurations and not in quantitative absolute values. 
This makes the exact values for the MFPTs arbitrary.
We however keep length dimensions in units of nm to better compare distances with the underlying biophysical system.

In Fig.~\ref{fig4} we show calculated MFPTs as a function
of the initial stagger $s_0$ for different NM2-variants in a semi-log plot for parallel and antiparallel configurations.
As expected, $\mathcal{T}_1$ decreases monotonically for increasing initial staggers $s_0$
as the overall distance to reach the absorbing target becomes smaller.
Surprisingly, however, this decrease occurs in only a few discrete drops, which
seem to correspond to specific barriers in the complex energy landscape. Moreover
these drops can be very large, with downward jumps of several orders of magnitude.
Together, these results imply that the MFPTs are both robust and very sensitive
to initial stagger values, depending whether one starts in one of the plateau
regions or close to an edge. Regarding the different isoforms, we see from Fig.~\ref{fig4} that
NM2A and NM2B behave very similarly, while NM2C is much less stable, with
the MFPTs being smaller by four or even six orders of magnitude for
parallel and antiparallel rods, respectively. This agrees with the fact 
that NM2C forms smaller minifilaments with approximately only
14 motor proteins compared to otherwise 30~\cite{dasbiswas2018ordering,craig2006structure}.
We also note that the difference between NM2A and NM2B seem to be a bit larger
for the antiparallel configurations. Interestingly, for parallel rod configuration, the four most
noticeably drops of the MFPT are approximately spaced with a 
$196$ residue repeat ($\sim\SI{28.54}{\nm}$), which is well known for
the charge repeat of NM2 molecules~\cite{mclachlan1982periodic,straussman2005skip, kaufmann2020electrostatic}.

It is clear that starting after a prominent barrier will result in a lower MFTP. 
However, the double-integral formula of the MFPT in Eq.~\eqref{eq:mftp_dimer_cont} not 
only takes local information ($s_0$) into account, but also encodes
global information of the energy landscape, because many different paths can
be taken towards the endpoint. 
This makes it notoriously difficult to answer the question which energy barriers 
are central to the shape of the MFPT in complex landscapes as in the case of our NM2-example.
Here we see that overcoming certain potential barriers substantially decreases the MFPT $\mathcal{T}_1$,
while others (possible even of same height) seemingly leave the MFPT unchanged. In order to better understand this
important point, we next turn to a coarse-graining procedure that asks the
question which features have to be kept when going to larger scales.

\begin{figure}[t]
	\includegraphics{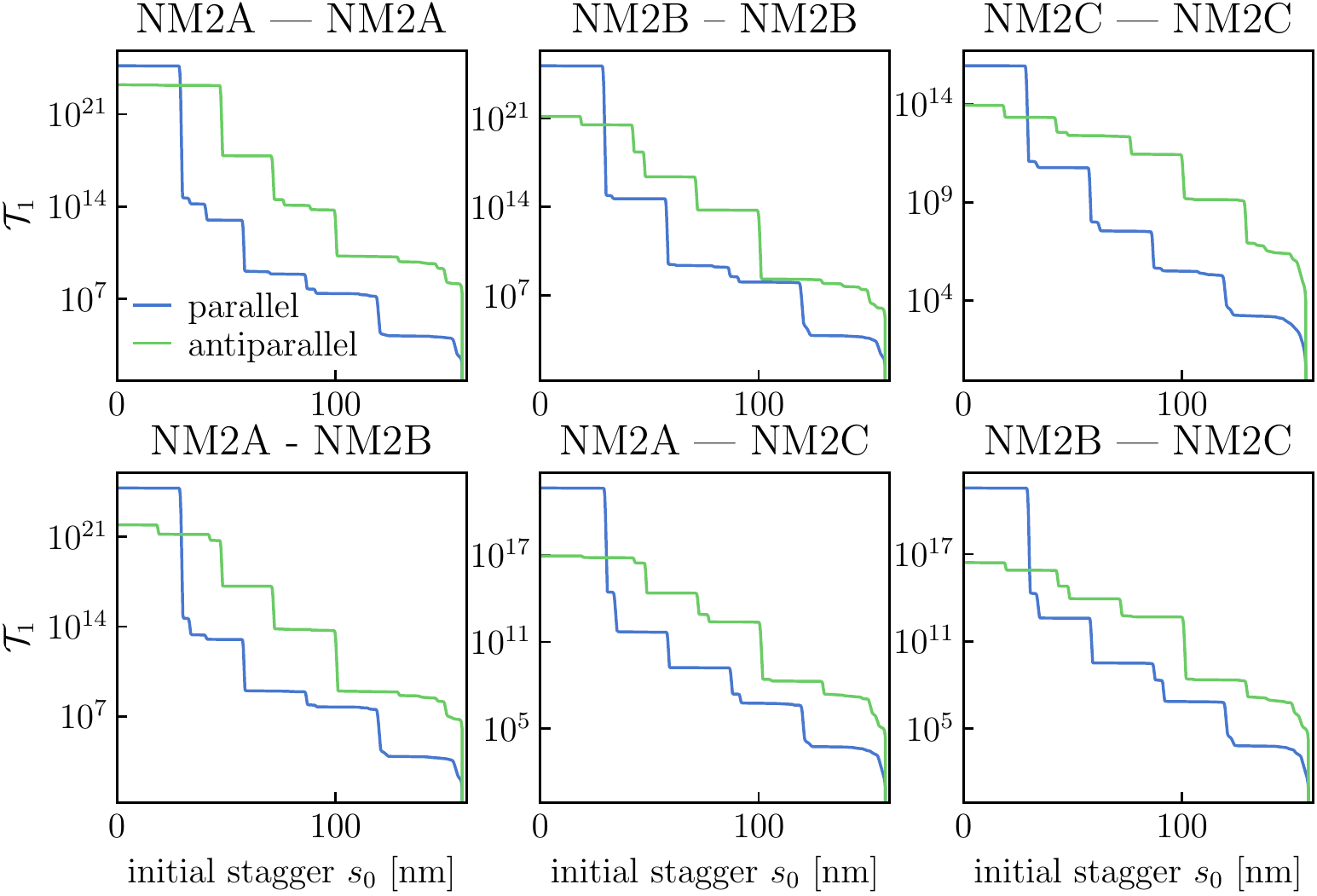}
	\caption{
	Direct comparison between the MFPT $\mathcal{T}_1$ for parallel (blue) and antiparallel (green) 
	NM2-dimer configurations for all different variant combinations.
	For the majority of initial staggers the antiparallel rod configuration has larger $\mathcal{T}_1$ 
	compared to parallel alignment, suggesting that antiparallel NM2-dimers are more stable.
	Only at small $s_0$ parallel NM2-rods have higher $\mathcal{T}_1$. All combinations
	involving NM2C have much smaller values for $\mathcal{T}_1$.}
    \label{fig5}
\end{figure}

In Fig.~\ref{fig5} we additionally report our results for all possible combinations of the three 
different isoforms. While the ordering does not matter here (different from our earlier
treatment that included bending~\cite{kaufmann2020electrostatic}), the alignment is very important.
We see that the antiparallel configurations (green) typically have higher values
for the MFPT $\mathcal{T}_1$ than the parallel ones (blue), indicating the 
importance of the bipolar architecture. Only for very small staggers
does the parallel configuration become more stable. Moreover we see again that 
combinations involving NM2C are always much less stable than the ones 
involving only NM2A and NM2B.

\section{Coarse-graining to master equation}

\subsection{One-step master equation}

Our numerical treatment of Eq.~\eqref{eq:mftp_dimer_cont}, which is based on 
continuous-state dynamics of an overdamped Fokker-Planck equation, has shown a pattern in the MFPT as
a function of stagger that requires an explanation in terms of the key features of the
underlying complex energy landscape. To identify these features, we 
now coarse-grain the continuous-state dynamics into a set of discrete states to further analyze the MFPT as a function of the initial stagger and learn more about the connection between energy landscape and the shape of the MFPT.
The motivation hereby is twofold.
First, the existence of high energy barriers that are significant to the thermal energy $k_{\rm B}T$
(see Fig.~\ref{fig3}) allows us to motivate a clear time-scale separation such that
a diffusing particle would spend the majority of 
its time located in valleys of the energy landscape before eventually escaping the barrier---a situation well described by stochastic transitions on a discrete network in terms of  
a Markovian jump process.
Second, the MFPT until a NM2-dimer detaches has, compared to the underlying energy landscape, a relatively simple shape suggesting that only a selected few potential barriers and minima contribute to the overall average time needed.
As seen in Fig.~\ref{fig4} the shape of the MFPT is characterized by a of number pronounced decreases which coincide with the position of specific potential barriers.
Unfortunately this feature remains illusive for the continuous-state description as
the expression used to calculate the MFTP $\mathcal{T}_1$ in Eq.~\eqref{eq:mftp_dimer_cont}
poses a non-trivial transformation of the underlying energy landscape.

\begin{figure}[t]
\includegraphics{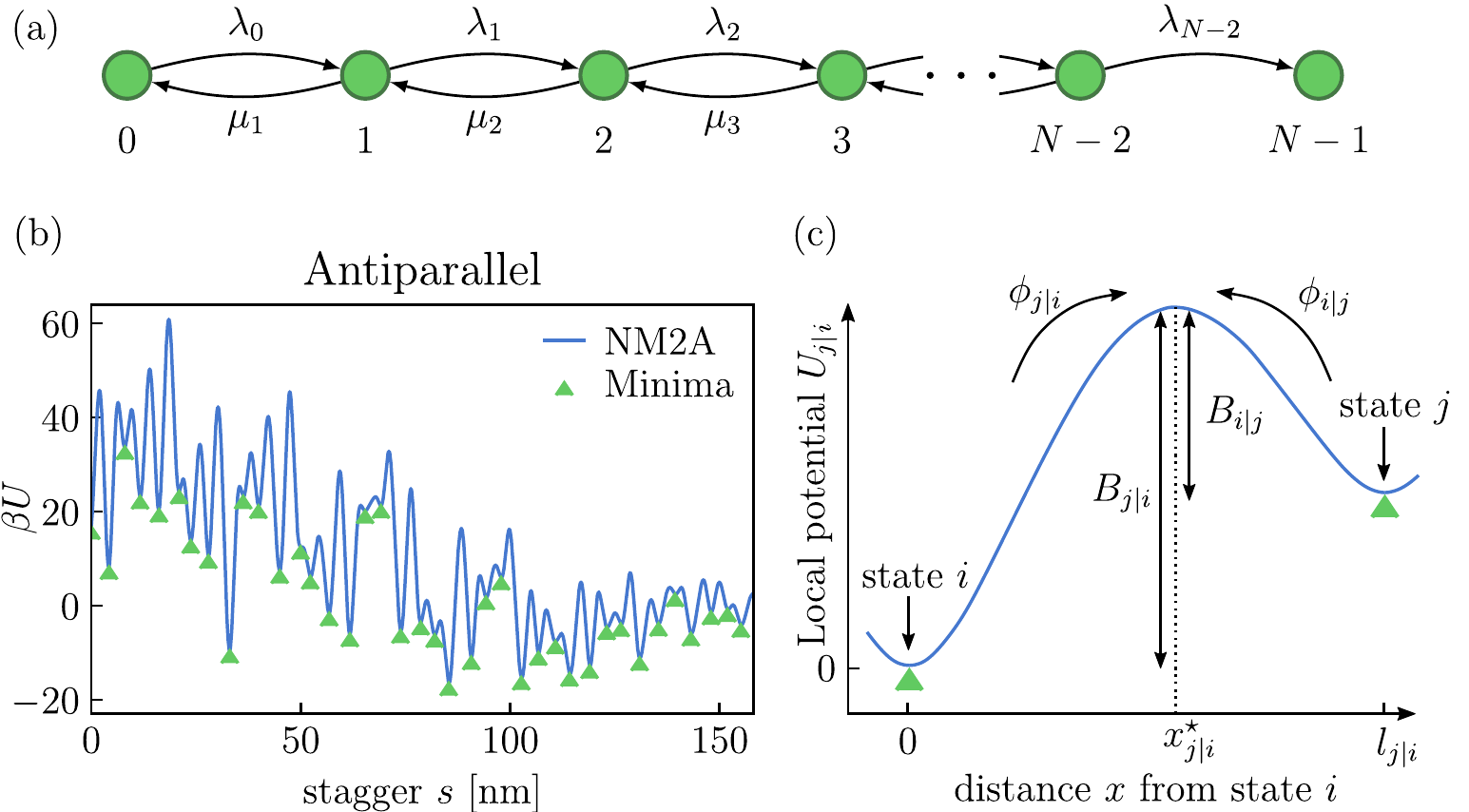}
\caption{Master equation. (a)
		Network of the coarse-grained Markov jump process describing the minima-to-minima dynamics of the original continuous-state dynamics. States correspond to the identified minimas and transitions rate $\lambda_i$ and $\mu_i$ are obtained as detailed in the main text.
		The last state $N-1$ describes the detachment of a NM2-dimer, i.e., it is absorbing.
        (b) Energy landscape of the antiparallel NM2A dimer without pulling force (i.e.~$F=\SI{0}{pN}$).
        Highlighted are all 39 local potential minima (green) each corresponding to a discrete state in the
        network description of the coarse-grained system.
		(c) Local energy landscape $U_{j|i}$ between adjacent states $i$ and $j$ which are separated by a 
		barrier located at position $x_{j|i}^\star$.
		If both barriers are high (e.g.~$B_{j|i}$ and $B_{i|j} > k_{\rm B}T$), the systems locally equilibrates prior to the next transition and can be described using a Markov jump process.	
		The splitting probability $\phi_{j|i}$ denotes the probability that the next transition $i\to j$ takes
		place.
		}
	\label{fig6}
\end{figure}

Within the coarse-graining procedure we treat each local minima of the corresponding NM2 energy landscapes
as a discrete state in the network description. 
Being in a state is therefore equivalent to residing in the associated local basin.
By splitting the energy landscape into a number of $N$ basins (compare Fig.~\ref{fig6}b),
delimited by their adjacent potential maxima, the resulting Markovian network is then described 
by the set $\Omega = \{0,1,\ldots, N-1\}$
of $N$ discrete states.
Local minima in the energy landscape are identified numerically using the \texttt{signal.argrelmin} function of the Python library \texttt{SciPy}~\cite{2020SciPy-NMeth}.
The network dynamics follow the master equation
\begin{equation}
	\partial_t p_i(t) = \sum_{j=1}^{N-1} [w_{ji} p_j(t) - w_{ij}p_i(t)].
	\label{eq:master_equation}
\end{equation}
Here $p_i(t)$ represents the occupation probability to be in state $i$ at time $t$ and $w_{ij}$ denotes the transition rate to pass from state $i$ to state $j$. 
Transitions only occur between directly neighboring states due to the considered problem
of stagging two neighboring rods. Therefore, the Master equation~\eqref{eq:master_equation} may be re-written in a one-step form as 
\begin{equation}
	\partial_t p_i(t) = \lambda_{i-1} p_{i-1}(t) + \mu_{i+1} p_{i+1}(t) - (\lambda_i + \mu_i) p_i(t)
	\label{eq:onestep_master}
\end{equation}
where we define the transition rates $\lambda_i = w_{i,i+1}$ and $\mu_i = w_{i,i-1}$
(transition rates between non-neighboring states vanish).
Fig.~\ref{fig6}a shows the corresponding reaction network. The MFPT is now identified with
the time at which the transition $N-2 \to N-1$ into the absorbing state takes place.
As before, the state $N = 0$ is reflecting.

\subsection{Transition rates}

Without loss of generality, we now choose the NM2A dimer in antiparallel configuration as an exemplary case.
Fig.~\ref{fig6}b shows the $N=39$ states that we obtain numerically from the corresponding energy landscape.
For Markovian network dynamics, the exit rate $r_i$ (or inverse lifetime) of the state $i$, i.e., the time that passes until the state changes, is exponentially distributed according to 
$p_i^{\rm exit}(t) = r_i \exp(-r_i t)$
with $r_i = 1 / \langle t \rangle_i^{\rm exit} = \sum_{j \in \mathcal{N}_i} w_{ij}$
(here $\mathcal{N}_i$ labels the neighbors of state $i$).
The exit time distribution is independent of the final state $j$, allowing us to write
the (splitting) probability of the transition from state $i$ to $j$ (i.e., that the next visited state will be $j$ conditioned on that the system is in state $i$) as 
$\phi_{j|i} = w_{ij} / r_i$.
Under these considerations the transition rates are then obtained as
\begin{equation}
	w_{ij} = \frac{\phi_{j|i}}{\langle t \rangle_i^{\rm exit}}.
	\label{eq:trans_rate}
\end{equation}
Quantities like the average exit time and the splitting probability entering Eq.~\eqref{eq:trans_rate} can be
related to a \emph{conditional} first-passage time problem~\cite{li_mechanisms_2013,hartich2021emergent}.
Here, one seeks the first time a system reaches a specific target state conditioned on that it has not yet reached any other (of possibly arbitrary many) target states before.
Based on this framework and using the renewal theorem~\cite{siegert1951first}, 
results for the splitting probability and the mean exit time have been derived recently
for general networks with a ``star-like'' topology~\cite{hartich2021emergent}.
The ``inner'' node is thereby the starting state and all ``outer'' nodes are target states. 
A line network as we consider it here is a particular simple case of this network topology.
Hereby, assuming that the full network dynamics between two network states evolve according to a continuous space-time overdamped Markovian diffusion,
the \emph{splitting probability} $\phi_{j|i}$ that starting from some state $i$ the neighboring 
state visited next will be state $j$ is given by the explicit expression
\begin{equation}
	\phi_{j|i} = \frac{1/ I_{j|i}^{(1)}}{\sum_{k \in \mathcal{N}_i} 1 / I^{(1)}_{k|i}}
	\label{eq:split_prob}
\end{equation}
and is normalized such that $\sum_{j \in \mathcal{N}_i} \phi_{j|i} = 1$.
The second important quantity required to obtain transition rates is the mean exit time of state $i$ which is obtained using the same framework via the expression
\begin{equation}
	\langle t \rangle_{i}^{\rm exit} = \sum_k \phi_{k|i} I_{k|i}^{(2)}
	\label{eq:exit_time}
\end{equation}
where the two auxiliary integrals $I^{(1)}_{j|i}$ and $I_{j|i}^{(2)}$ in Eqs.~\eqref{eq:split_prob} and~\eqref{eq:exit_time} are defined~\cite{hartich2021emergent} as

\begin{equation}
	I_{j|i}^{(1)} = \int_0^{l_{j|i}} dy_1 \ g_{j|i}^{(1)} 
	\qquad \textrm{with} \qquad 
	g_{j|i}^{(1)} = \frac{1}{D_{j|i}} e^{\beta U_{j|i}(y_1)}\ ,
	\label{eq:int1}
\end{equation}

\begin{equation}	
	I_{j|i}^{(2)} = \int_0^{l_{j|i}} dy_1 \int_0^{y_2} dy_2 \ g_{j|i}^{(2)}  
	\qquad \textrm{with} \qquad g_{j|i}^{(2)} 
	= \frac{1}{D_{j|i}} e^{\beta U_{j|i}(y_1)- \beta U_{j|i}(y_2)}\ .
	\label{eq:int2}
\end{equation}
Here $l_{j|i}$ denotes the distance between two adjacent minima (i.e.~states $i$ and $j$) defined
through the underlying energy landscape and 
$U_{j|i}$ is the \emph{local} energy landscape relative to the value at state $i$ between states $i$ and $j$ as 
schematically depicted in Fig.~\ref{fig6}.
For the purpose of calculating the transitions rate between two neighboring states we therefore identify their intervening local potentials and numerically compute the integrals in Eqs.~\eqref{eq:int1} and~\eqref{eq:int2} employing the trapezoid rule function 
\texttt{integrate.trapezoid} as implemented
in the python library \texttt{SciPy}~\cite{2020SciPy-NMeth}.

\subsection{Mean first-passage time}

\begin{figure}[t]
	\includegraphics[width=0.9\textwidth]{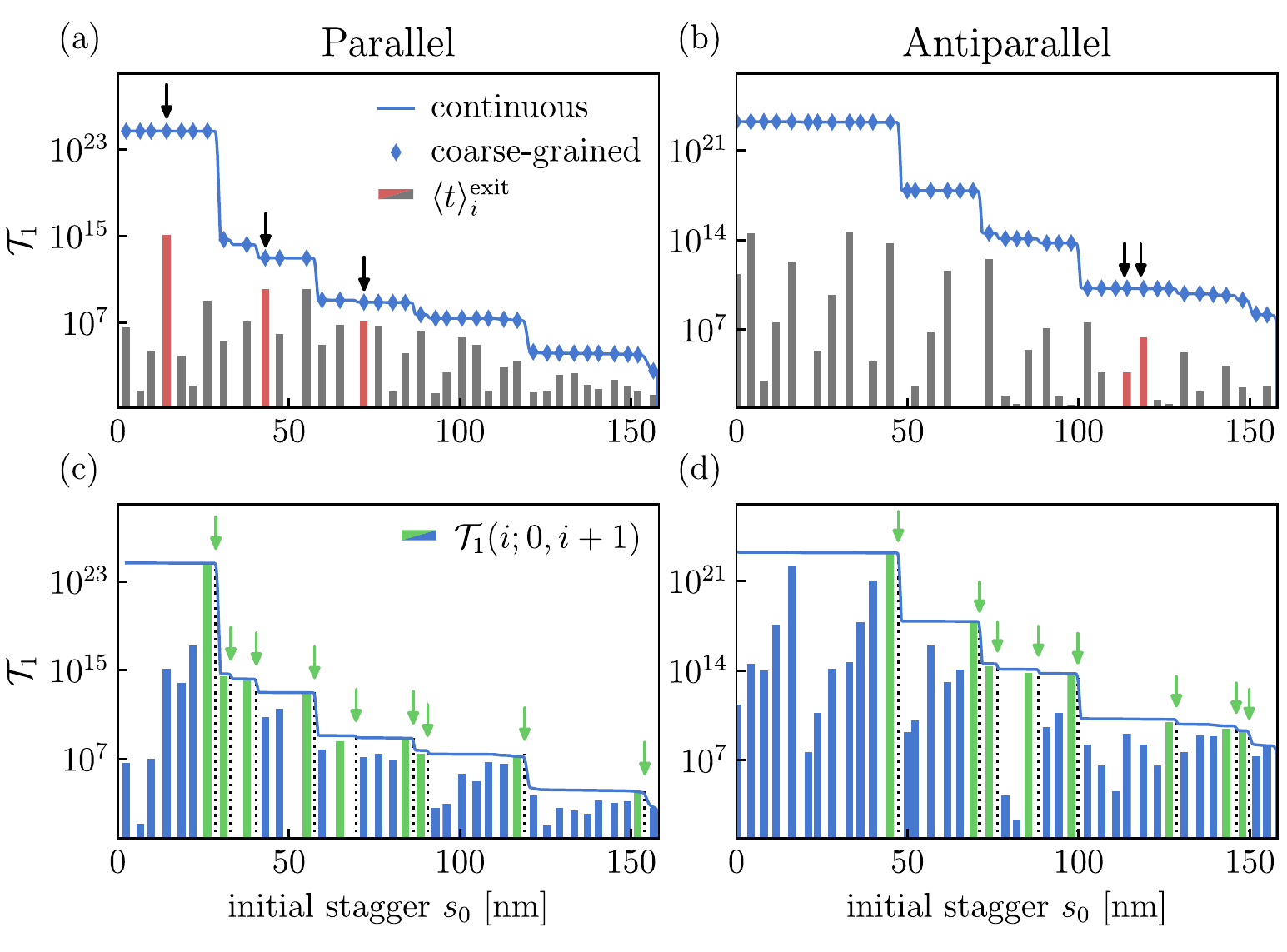}
	\caption{
	Mean first-passage times (MFPT) $\mathcal{T}_1$ for the original continuum system 
	compared to its coarse-grained counterpart.
	(a,b) Comparison between the MFPT $\mathcal{T}_1$ of the continuous description Eq.~\eqref{eq:mftp_dimer_cont}
	and the coarse-grained network Eq.~\eqref{eq:mftp_network} for (a) parallel and (b) antiparallel 
	NM2A-dimer configurations.
	Mean exit-times $\langle t \rangle_i^{\rm exit}$ (Eq.~\eqref{eq:exit_time}) are depicted as bars.
	Experimental staggers (black arrows) are not located at positions of large drops
	and are characterized by comparatively larger mean exit-times to their neighbors (red bars), ensuring local robustness.
	(c,d) Single MFPTs $\mathcal{T}_1(i;0,i+1)$ from state $i$ to $i+1$ are denoted as bars for the (c) parallel and (d) antiparallel cases.
	Largest contributions that give rise to jumps in the shape of the overall MFPT are highlighted in green.
	Predictions of the large jumps obtained using the network description as detailed in the main text are highlighted with green arrows and dotted lines and agree very well with the drops in the full solution.}
	\label{fig7}
\end{figure}

After identifying all network states and calculating the corresponding transition rates $w_{ij}$,
and thus $\lambda_i$ and $\mu_i$ of Eq.~\eqref{eq:onestep_master},
the Markov jump process is now fully specified.
Therefore, we are able to determine the mean first-passage time (MFPT), i.e.~the average detachment time of NM2-dimers, within the coarse-grained network description
of the minima-to-minima dynamics of a diffusing Brownian particle in the original overdamped Fokker-Planck picture.
As for the continuous system, the corresponding absorbing state is hereby always placed in the last state such that reaching it corresponds to the detachment of the NM2-molecules.

In analogy to the analytical solution of the MFPT in Eq.~\eqref{eq:mftp_dimer_cont}, which holds for a Fokker-Planck equation for overdamped Langevin dynamics in an external potential $U$,
for a Markov process confined between mixed boundaries ($a$ reflecting and $b$ absorbing with $a < b$)
the MFPT is given by~\cite{li_mechanisms_2013,goel2016stochastic}
\begin{equation}
	\mathcal{T}_1(i_0;a,b) = \sum_{i=i_0}^{b-1} \sum_{n=a}^{i} \lambda_n^{-1} \Pi_{n+1,i} \quad \textrm{with} \quad 
	\Pi_{i,j} = \frac{\mu_i \mu_{i+1} \cdots \mu_j}{\lambda_i \lambda_{i+1} \cdots \lambda_j} \quad \textrm{for} \quad i \leq j,
	\label{eq:mftp_network}
\end{equation}
where $i_0$ is the starting state, $a$ the reflecting state, and $b$ the absorbing target state.
In our case this MFPT has to be calculated as a function of the initial stagger $i_0$
and $a$ and $b$ are set to the first and last state, respectively.
This calculation now is much more computationally efficient than the continuum version.
Figs.~\ref{fig7}a and b show a direct comparison between the MFPT $\mathcal{T}_1$ computed using Eq.~\eqref{eq:mftp_dimer_cont} (lines) and Eq.~\eqref{eq:mftp_network} (symbols) for the NM2A-dimer.
Clearly, the results of the two different approaches agree very well with each other,
thus validating our coarse-graining approach. 
The same picture emerges for the other NM2 variants.
We additionally plot the mean exit-time $\langle t \rangle_i^{\rm exit}$ from Eq.~\eqref{eq:exit_time} as bars.
Interestingly, the experimentally observed staggers (black arrows) for the antiparallel configuration corresponds to large values for $\langle t \rangle_i^{\rm exit}$. In particular, the most prominent stagger at $s=\SI{14.3}{\nm}$ corresponds to the highest overall mean exit-time.
While the mean exit-times corresponding to experimentally observed staggers for antiparallel alignment are not notably large on a global scale, they are indeed large compared to adjacent states.
This suggests that these staggers confer some local robustness against perturbations.
We additionally note that the periodicity between small mean exit-times seems to follow the characteristic $98$-residue repeat pattern already encountered earlier.

\subsection{Predicting jumps}

The relatively simple shape of the MFPT and the characteristic feature to decrease substantially over several orders of magnitude by only a slight variation of the initial stagger position between two NM2-molecules in the detachment process remained elusive in the continuous description. The discrete nature of the Markov jump process
now allows us to gain further insights in the behavior of the MFPT of the detachment process of two aligned NM2-molecules.
We first note that an exploration of the complete network topology (i.e.~the complete energy landscape) takes place if the system is initially prepared in state $i_0 = 0$. Then is has to visit each state at least once before reaching the absorbing target at the other end of the network chain.
For this case Eq.~\eqref{eq:mftp_network} is explicitly given by
\begin{equation}
\label{eq:mfpt_markov_dimer}
    \begin{aligned}[b]
	\mathcal{T}_1(0;0,b)
	&= \sum_{i=0}^{b-1} \sum_{n=0}^{i} \lambda_n^{-1} \Pi_{n+1, i}
	\\
	&= 
	\underbrace{\lambda_0^{-1}}_{i=0}
	+
	\underbrace{\lambda_0^{-1} \Pi_{1,1} + \lambda_1^{-1}}_{i=1}
	+
	\underbrace{\lambda_0^{-1} \Pi_{1,2} + \lambda_1^{-1} \Pi_{2,2} + \lambda_2^{-1}}_{i=2}
	\\	
	&+
	\ldots
	+ \underbrace{\lambda_0^{-1} \Pi_{1,b-1} + \lambda_1^{-1} \Pi_{2,b-1} + \ldots +  \lambda_{b-2}^{-1} \Pi_{b-1,b-1} + \lambda_{b-1}^{-1} }_{i=b-1}.
\end{aligned}
\end{equation}
For a process that starts at a transient initial state $m$ with $0 < m < b$, i.e.~away from the reflecting state, the lower bound of the first sum in Eq.~\eqref{eq:mfpt_markov_dimer} has to be changed from $i=0$ to $i=m$.
Explicitly writing out the double sum in this way shows that terms of the outer sum (summation over $i$) correspond to \emph{single} MFPTs $\mathcal{T}_1(i;0,i+1)$ from state $i$ to $i+1$ while keeping the reflective state unaltered in state $0$.
That is, irrespective of where the systems starts, it is always allowed to visit state ``0'' again.
For illustration, consider the single MFPT $\mathcal{T}_1(2;0,3)$ for the transition $2 \to 3$, i.e.~the case $i=2$ above.
The last contribution, $\lambda_2^{-1}$, is the time needed for the direct path from  state $2$ to $3$. 
Depending on the transitions rates (e.g.~when $\mu_3 > \lambda_2$) the system might however also transition backwards to state $1$ or even $0$ before finally reaching the absorbing target in state $3$ for the first time.
These additional paths are accounted for by the corresponding inverse rates and the weighting factor 
$\Pi_{2,2}$ as defined
in Eq.~\eqref{eq:mftp_network}.
Consequently, we may express the MFPT as a sum over all individual single MFPTs as
\begin{equation}
	\mathcal{T}_1(0;0,b) = \sum_{i=0}^{b-1} \mathcal{T}_1 (i; 0, i+1)
	= \mathcal{T}_1 (0;0,1) + \mathcal{T}_1 (1;0,2) + \ldots + \mathcal{T}_1 (b-1;0,b),
	\label{eq:single_mfpt}
\end{equation}
where the systems starts initially in state 0.
In the same way as before, if one starts in a state $0 < m < b$ the summation starts from $i=m$.

A decomposition into single MFPTs now makes it possible to distinguish between larger and smaller contributions to the full MFPT. 
Single MFPTs from state $i$ to state $i+1$ (leaving the reflecting state 0 unchanged)
are shown in Fig.~\ref{fig7}c and d for the network description of the NM2-dimer system for antiparallel and parallel
configuration, respectively. It becomes immediately evident that the single MFPTs are highly irregular and vary
over several orders of magnitude depending on the initial state $i_0$.
Based on this coarse-grained description we are now able to identify the largest contributions towards the average detachment time of the NM2 dimer (marked in green).
This implies that if we were to initially prepare the system in state $m > i^\star$, where $i^\star$ denotes the state with the largest single MFPT $\mathcal{T}_1(i;0,i+1)$, we would expect 
a substantial decrease in the time needed until the NM2-dimer detaches as all
contributions (and therefore the largest too) of single MFPTs for states $i^\star < m$ drop out, i.e., 
the sum in Eq.~\eqref{eq:single_mfpt} starts at $i=m$.
Analogously, the next and all further drops in the MFPT can be identified using the same principle,
namely finding the next largest single MFPT and choosing the initial state accordingly.
In Fig.~\ref{fig7}c and d we show that indeed our procedure of
identifying the relevant minima and barriers (green bars) predicts the observed drops in the full solutions (green arrows and dotted lines), both for
parallel (c) and antiparallel (d) configurations. Thus our main
question has now been answered, namely which energy barriers
lead to the large drops in the MFPTs.

\section{Mean first-passage time under mechanical load}
\label{sec:MFPT_Force}

Finally we address the question how 
the MFPT changes if we ``tilt'' the energy landscape by an additional pulling force as motivated in Sec.~\ref{sec:elec_int}.
More precisely, in our biophysical example of the NM2 assembly landscape, we now study how
the average time until NM2-dimer detachment behaves under a variable mechanical load
that is applied along the dimer main axis.

\begin{figure}[t]
	\includegraphics{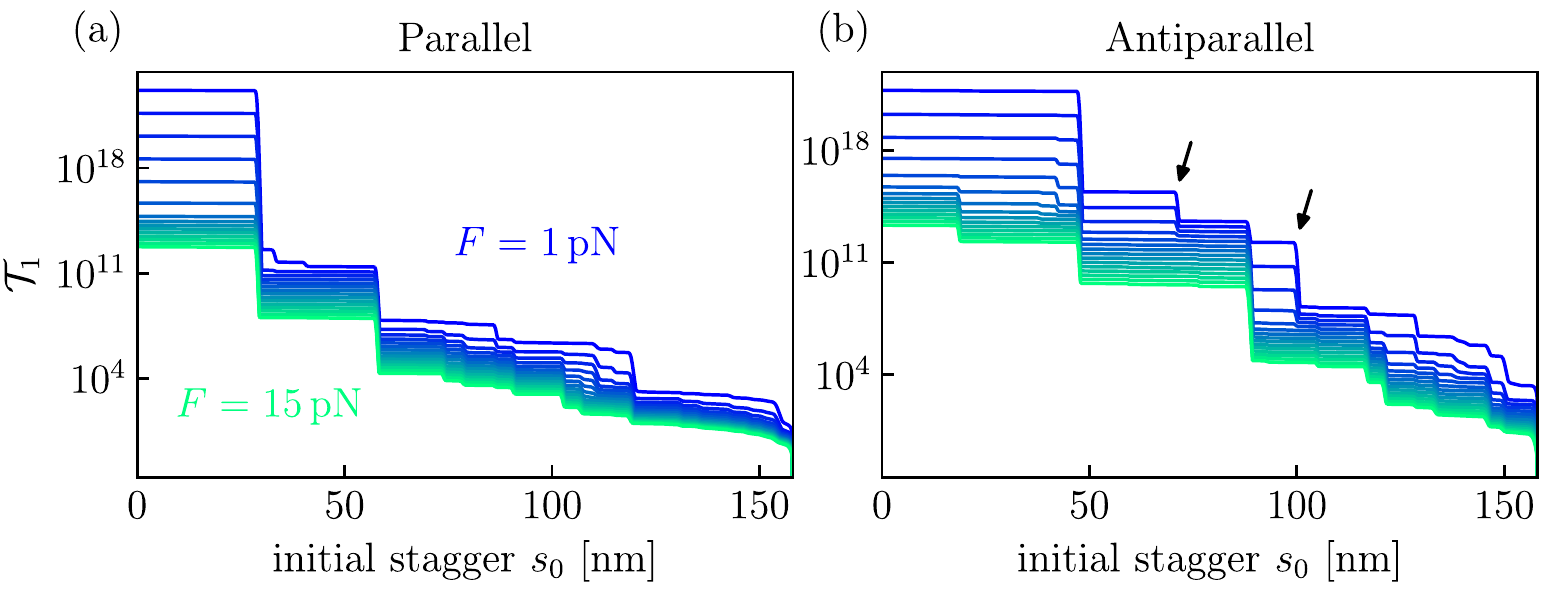}
	\caption{Mean first-passage time (MFPT) $\mathcal{T}_1$ as a function of the initial staggering $s_0$ between two aligned NM2-molecules for variable pulling forces $F$ ranging from $F=\SI{1}{pN}$ (blue) to $F=\SI{15}{pN}$ (teal) in steps of one.
	Depicted are the MFPTs for the NM2A variant in (a) parallel and (b) antiparallel configurations.
	While the shape for parallel configuration does not change much, the antiparallel dimer shows 
	qualitative changes as sudden changes in $\mathcal{T}_1$ vanish by increasing the pulling forces,
	at positions highlighted by the black arrows.}
	\label{fig8}
\end{figure}

To compute the MFPT for the driven system we insert the tilted energy landscape 
Eq.~\eqref{eq:tilted_landscape} into Eq.~\eqref{eq:mftp_dimer_cont}.
Resulting MFPTs of antiparallel and parallel alignments are depicted in 
Fig.~\ref{fig8} for different pulling forces $F$.
As one would expect, the application of the pulling force results in faster MFPTs to reach the absorbing border, i.e., faster detachment of aligned NM2-molecules. 
Interestingly, the MFPT for the parallel NM2-dimer mostly keeps its qualitative shape unchanged.
This means that the set of important potential barrier that lead to large decreases in the MPFT
remains the same even under the presence of external pulling.
In contrast, shapes of the MFPT for antiparallel orientation undergo more pronounced changes under the influence of external pulling forces.
On one hand, overcoming previously identified significant potential barriers at e.g.~$\approx\SI{72}{\nm}$
or $\approx \SI{98}{\nm}$ (black arrows) now does not result in a large decrease of the MFPT anymore.
On the other hand, overcoming new potential barriers at e.g.~$\approx \SI{87}{\nm}$ or
$\approx \SI{117}{\nm}$ now results in a large decrease of the MFPT.
We suggest that this apparent difference between antiparallel and parallel alignment might
have biophysical consequences during NM2 minifilament assembly and filament splitting during
partition events under mechanical load.

\section{Conclusions}

Here we have studied the mean first-passage time (MFPT) for rough, complex one-dimensional energy landscapes.
As a biophysical relevant example we considered the self-assembly process of the
three different isoforms A, B and C of nonmuscle myosin II (NM2).
We theoretically studied the average time until two aligned NM2-molecules (i.e.~the nucleation point for minifilament assembly) detach with and without an additional pulling force as a first-passage time problem on complex energy landscapes derived from the amino acid sequences of different NM2-variants.
We demonstrated how the MFPT of highly complex and rugged one-dimensional landscapes can be obtained by coarse-graining the underlying continuous dynamics described by the Fokker-Planck equation into a discrete-state continuous-time Markov jump process, that is a master equation. Our numerical procedure not only gives perfect agreement 
between the two descriptions, it also allowed us to decompose the MFPT into its different contributions
and from there to identify the relevant barriers that lead to the sudden drops in lifetime. Overall, our
work shows how one can better understand the relevant features of a complex energy
landscape by systematically studying the corresponding MFPTs.

Because here we restrict ourselves to the one-dimensional case
of straight rods, our MFPT-treatment does not predict the experimentally
observed staggers as well as did our earlier treatment that included the bending
energy of the rods splaying away from the main axis due to electrostatic
repulsion \cite{kaufmann2020electrostatic}. Nevertheless
our specific results for NM2 also seem to be very important to better understand the biophysics of NM2-
minifilaments and in particular shed new light on the role
of physical force for their assembly and stability. In detail, we found that
the experimentally known staggers are in regions of high stability (away from the 
jumps), that they exhibit large local exit times,
that dimers involing NM2C are much less stable (shorter lifetimes) than the ones 
involving NM2A and NM2B, that the difference between NM2A and NM2B is larger for
the antiparallel configurations, that antiparallel configurations are more
stable without force, but also show larger changes under pulling force. In the future, 
it would be very interesting to extend this approach to the 
energy landscapes including bending as well as to go beyond dimers and to 
also consider high-order assemblies. This, however, would require to treat
high-dimensional energy landscapes which constitutes formidable numerical
and mathematical challenges.

\section*{Acknowledgments}

We acknowledge funding through the Max Planck School Matter to Life, 
supported by the German Federal Ministry of Education and Research (BMBF) in collaboration with the Max Planck Society. 
We thank Tom Kaufmann, Justin Grewe, Kai Weißenbruch, Martin Bastmeyer and Jordan Beach for stimulating discussions.


\end{document}